\documentstyle[preprint,aps]{revtex}
\newcommand{\half}{\frac{1}{2}}
\newcommand{\bea}{\begin{eqnarray}}
\newcommand{\eea}{\end{eqnarray}}
\newcommand{\be}{\begin{equation}}
\newcommand{\ee}{\end{equation}}
\begin{document}
\draft
\title{Projectile $\Delta$ Excitations in
$p(p,n)N\pi $ Reactions}
\author{Yung Jo and  Chang-Yong Lee}
\address{Department of Physics, University of Texas, Austin, Texas 78712}
\date{June 23, 1995}
\maketitle
\begin{abstract}
It has recently been proven from measurements of
the spin-transfer coefficients $D_{xx}$ and $D_{zz}$ that
there is a small but non-vanishing $\Delta S=0$ component
$\sigma_{0}$, in the inclusive $p(p,n)N\pi\,$
reaction cross section $\sigma\,$.
It is shown that the dominant part of the measured $\sigma_{0}$
can be explained in terms of the projectile $\Delta$ excitation
mechanism. An estimate is further made of
contributions to $\sigma_{0}$ from  s-wave rescattering
process. It is found that s-wave rescattering contribution is much
smaller than the contribution
coming from projectile $\Delta$ excitation mechanism.
The addition of  s-wave rescattering contribution to the dominant
part, however, improves the fit to the data.
\end{abstract}

\vspace{2ex}


The $p(p,n)N\pi$
reaction at intermediate energies
has been a subject of a number of studies from both
experimental~\cite{Glas1,Glas2,Shim,Wick,Prou} and
theoretical~\cite{Jain,Silb,Mizu,Ray} point of views.
The understanding of the
reaction is important in its own sake; it is one of the basic
processes in the intermediate nuclear physics.
One of the dynamical process involved in the reaction
is the projectile $\Delta$ excitation process (PDP). The PDP
is usually ignored in the inclusive $(p,n)$ cross section $\sigma$
calculations, since $\sigma$ is dominated by the contribution
coming from the target $\Delta$ excitation process (TDP). The
contribution from PDP gives only a small correction
to the dominant TDP cross section. Therefore, it has been difficult
to test the predicted PDP cross section by the inclusive cross section
data.

Recently, however, several measurements of the
spin-transfer coefficients $D_{xx}$ and $D_{zz}$ have been
made~\cite{Glas1,Glas2,Prou}. Using these coefficients,
it is possible to extract the no spin-transfer($\Delta S=0$) component
$\sigma_{0}$ from the inclusive cross section $\sigma\,$.
In fact, we show that the measured $\sigma_{0}$ can be explained well
in terms of PDP.
 In the present study, we restrict our interests to
the zero-degree case, i.e., the case where the neutron is
emitted at zero-degree ($\theta_{n}=0^{o}$).  Under this restriction,
$\sigma_{0}$ can be expressed, in terms of the observed inclusive
cross section $\sigma$ and the spin-transfer
coefficients $D_{xx}$ and $D_{zz}\,$, as
\begin{equation}
\sigma_{0}=\frac{1}{4} \sigma\, (1+2D_{xx}+D_{zz})\,.
\label{eq1}
\end{equation}

\vspace{0.1in}
 In order to present the theoretical cross sections
$\sigma_{0}$ and $\sigma$, let us denote the
$p(p,n)N\pi$ reaction as
$a +A \longrightarrow b + B + \pi^{\alpha}$,
where $a$ ($b$) and $A$ ($B$)
denote the projectile (ejectile) and target (residual nucleus)
respectively, and $\pi^{\alpha}$ is the emitted pion that carries
the charge $\alpha\,$.
In the center of mass system, total inclusive cross section $\sigma$
may be written as
\begin{eqnarray}
\sigma = \left. \frac{d^{2}\sigma}{dE_{b} d\Omega_{b}} \right|_{\theta_b =0} =
\frac{m_{a} m_{b} m_{A} m_{B}}{(2\pi)^{5}\, 2\,\sqrt{s}}
\frac{p_{b}}{p_{a}} \int d\Omega^{d}_{\pi} \,
\frac{p^{d}_{\pi}}{s_{d}}\,\overline{|T|^{2}}\, ,
\label{incl}
\end{eqnarray}
where $m_{i}$ and $p_{i}$ ($i=a,b,A,B$) are the mass
and the 4-momentum of the particle $i$,
$\;s_{d}$ is the invariant mass of the final $N+\pi$ system
and $s=(p_{a}+p_{A})^{2}$.
Further, $\Omega^{d}_{\pi}$ and $p^{d}_{\pi}$ are the solid angle and
the momentum of the emitted pion in the $N+\pi$ rest frame,
while $T$ is the Lorentz invariant transition amplitude.
$\overline{|T|^{2}}$ means to take the sum and average of the final and
initial spin states respectively.

\vspace{0.1in}
 In calculating $\sigma$, we take into account
both PDP and TDP as well as contributions from the s-wave
rescattering processes. These
processes are schematically shown in Fig.~1(a) and 1(b) (target and
projectile $\Delta$ excitations) and Fig.~1(c) and (d) (s-wave
rescattering processes).

First, the $\Delta$ excitation processes(for both PDP and TDP  in Figs.~1(a)
and 1(b) respectively) are treated by means of the transition amplitude
$\hat{t}_{NN,N\Delta}$ used in Ref.~\cite{Udag} and $\Delta$ decay
Hamiltonian.  The explicit form of $\hat{t}_{NN,N\Delta}$ is
\begin{eqnarray}
\hat{t}_{NN,N\Delta}=V_{L}(\hat{q} \cdot \vec\sigma)(\hat{q} \cdot
{\vec S}^{\dagger})
+V_{T}(\hat{q} \times \vec\sigma) \cdot (\hat{q} \times {\vec
S}^{\dagger}) \, ,
\label{tamp}
\end{eqnarray}
where $\hat{q}$ is the unit vector whose direction is that of the
momentum transfer involved in the excitation process, $\vec\sigma$ is the
Pauli spin operator and ${\vec S}^{\dagger}$ is the spin operator for the
$N \rightarrow \Delta$ transition.  $V_{L}$ and $V_{T}$ are strength
parameters of  the spin-longitudinal(LO) and spin-transverse(TR) which
are used in Ref.~\cite{Udag}.
The Hamiltonian for the $\Delta$ decay is
\begin{eqnarray}
H_{{\pi}N\Delta}=\frac{f^{*}}{\mu}\,({\vec p}_{\pi} \cdot {\vec
S}^{\dagger})\,T^{\alpha} \, + \,\mbox{ h.c. }\; ,
\label{eq4}
\end{eqnarray}
where $\mu$ denotes pion mass and $T^{\alpha}$ is isospin transition
operator with charge $\alpha\,$. For the coupling constant we take
${f^{\ast^{2}}}\!/{4\pi}=0.36\,$.

Second, the s-wave rescattering processes are calculated as in
Ref.~\cite{Oset}. The basic couplings in this process are $NN\pi$
coupling and $N\pi\longrightarrow N\pi$ s-wave amplitude. The $NN\pi$
coupling is given by
\be
 H_{\pi NN}=\frac{f}{\mu}\,({\vec p}_{\pi} \cdot
{\vec\sigma})\,\tau^{\alpha}\; ,
\ee
where ${\vec p}_{\pi} $ is the momentum of the pion and the coupling
is given as $f^{2}/4\pi=0.08\,$.
The Hamiltonian for the s-wave $N\pi\longrightarrow N\pi$ is given as
\be
H_{\pi \pi NN}=4\pi \delta_{m_{s}m'_{s}} \left\{ \frac{2\lambda_{1}}{\mu}
\delta_{m_{t}m'_{t}}\delta_{\lambda\lambda^{'}}
+i\epsilon_{\alpha\lambda\lambda^{'}}\frac{2\lambda_{2}}{\mu}
<m^{'}_{t}|\tau^{\alpha}|m_{t}> \right\}\, ,
\label{eqswave}
\ee
where indices $m_{s},m'_{s},m_{t},m'_{t}\,$ in Eq.~\ref{eqswave} are
the spin and isospin variables of the incoming and outgoing nucleons.
For the couplings, we take~\cite{Oset}
\be
\lambda_1\,=\,0.0075\;\;,\;\; \lambda_{2}\,=\,0.0528 \, .
\ee

\vspace{0.1in}
The total T-amplitude can then be given as
\begin{eqnarray}
-iT & = & \sum_{s_1 \mu_1 s_2 \mu_2}
(-1)^{\half -m_a}\,<\half ,m_b ;  \half ,-m_a \, |\,s_1 , \mu_1>
\nonumber \\
& & \hspace{0.5in}\times \,(-1)^{\half-m_A}\, <\half , m_B ; \half ,
-m_A \,|\,s_2 , \mu_2> \, C_{s_1 \mu_1 s_2 \mu_2}\, ,
\label{eq7}
\end{eqnarray}
where ($s_{1} , \mu_{1}$) and ($s_{2} , \mu_{2}$)
represent the spin transfers involved in the
$a \longrightarrow b$ and $A \longrightarrow B$
transition processes respectively.  The partial amplitude
$C_{s_1\mu_1 s_2 \mu_2} $
may be decomposed into the two contributions
$A_{s_1 \mu_1 s_2 \mu_2}$ and $B_{s_1 \mu_1 s_2 \mu_2}\,$, coming from the
$\Delta$ excitation and s-wave rescattering processes respectively :
\begin{eqnarray}
C_{s_1 \mu_1 s_2 \mu_2}=A_{s_1 \mu_1 s_2 \mu_2}+B_{s_1 \mu_1 s_2
\mu_2}\, ,
\label{eq8}
\end{eqnarray}
where
\begin{eqnarray}
& & A_{0000}=0.0 \, , \\
& & A_{1\mu 00}=-\frac{4}{3} \frac{f}{\mu}
\,[\,(\hat{q} \cdot {\vec p}_{\pi}) \hat{q}^{*}_{\mu}V_{L} +
({\vec p}^{*}_{\pi \mu}-(\hat{q} \cdot
{\vec p}_{\pi})\hat{q}^{*}_{\mu})V_{T}]\,G_{t}\,C_{t}\, , \\
& & A_{001\mu}=\frac{4}{3} \frac{f}{\mu}
\,[\,(\hat{q}' \cdot {\vec p}'_{\pi}) \hat{q}'^{*}_{\mu}V_{L} +
({\vec p}'^{*}_{\pi \mu}-(\hat{q}' \cdot
{\vec p}'_{\pi})\hat{q}'^{*}_{\mu})V_{T}]\,G_{p}\,C_{p} \, , \\
& & A_{1\mu_{1} 1\mu_{2}}=-\frac{2 \sqrt{2}}{3} \frac{f}{\mu}
\,[ \hat{q}^{*}_{\mu_{1}}(\hat{q} \cdot {\vec p}_{\pi})^{*}_{\mu_{2}}
(V_{L}-V_{T}) \,G_{t}\,C_{t}  \nonumber  \\
& & \hspace{3.5cm} + \,(-1)^{\mu_{1}}<1,-\!\mu_{1} ;  1 , \nu_{2}\,|\,1 ,
\mu_{2}>\,p^{*}_{\pi \nu_{2}}
V_{T}\,G_{t}\,C_{t}
\nonumber \\
& & \hspace{3.5cm} - \, \hat{q}'^{*}_{\mu_{2}}(\hat{q}' \cdot {\vec
p}'_{\pi})^{*}_{\mu_{1}}
(V_{L}-V_{T})\, G_{p}\,C_{p} \nonumber \\
& & \hspace{3.5cm} +\,  (-1)^{\mu_{2}}<1 , -\!\mu_{2} ; 1 , \nu_{2}\,|\,1
, \mu_{1}>\, p'^{*}_{\pi \nu_{2}}
V_{T}\,G_{p}\,C_{p}\,] \, \\
\mbox{and} & & \\
& & B_{s_1 \mu_1 s_2 \mu_2}= 8\pi \frac{f}{\mu}
\,[-\delta_{s_{1}1}\delta_{s_{2}0}\,\hat{q}^{*}_{\mu_{1}}
\sqrt{2}\,\lambda_{+} \, +\,
\delta_{s_{1}0}\delta_{s_{2}1}\,\hat{q}'^{*}_{\mu_{2}}\lambda_{0}\,]\,
D_{\pi} \,F_{\pi}\, .
\end{eqnarray}
In the above expressions,  $\,\lambda_{0}\,$ and $
\,\lambda_{+}\,$ are given as
\be
\lambda_{0}\,=\,-\frac{2\sqrt{2}}{\mu}\,\lambda_{2}\;\;\;,\;\;\;
\lambda_{+}\,=\, \frac{2}{\mu}\,(\lambda_{1}+\lambda_{2})\; ,
\ee
$C_{i}$ is  the isospin factor for $\pi N \Delta$ vertex and index $i$
refers both target($t$) and projectile($p$) $\Delta$ excitations.
The propagators and the pion form factor are defined as follows.
\bea
& & G_{i}=\frac{1}{ \sqrt{s_{i}}-M_{\Delta}+i\Gamma(s_{i})/2 }
\hspace{0.5in};\hspace{0.5in} \mbox{$\Delta$ propagator}\, , \\
& & D_{\pi}=\frac{1}{\omega^{2}-q^{2}-\mu^{2}}
\hspace{0.5in};\hspace{0.5in} \mbox{$\pi$ propagator}\, , \\
& & F_{\pi}=\frac{\Lambda^{2}-m_{\pi}^{2}}{\Lambda^{2}-t}
\hspace{0.5in};\hspace{0.5in} \mbox{$\pi$NN form factor with}\;
\Lambda = 1200 \,\mbox{MeV}.
\eea
It is then easy to see that
\begin{eqnarray}
\overline{|T|^{2}}=\frac{1}{4} \sum_{s_{1} \mu_{1} s_{2} \mu_{2}}
|C_{s_{1} \mu_{1} s_{2} \mu_{2}}|^{2}\, .
\label{eq14}
\end{eqnarray}
We note further that $\sigma_{0}$ can be evaluated by simply
picking up the component with $(s_{1}, \mu_{1})=(0,0)$, which
comes from both PDP and s-wave rescattering from the
projectile. Thus, defining $\overline{|T_{0}|^{2}}$ as
\begin{eqnarray}
\overline{|T_{0}|^{2}}=\frac{1}{4} \sum_{s_{2} \mu_{2}}
|C_{00 s_{2} \mu_{2}}|^{2}\, ,
\label{eq15}
\end{eqnarray}
$\sigma_{0}$ can be given as
\begin{eqnarray}
\sigma_{0} = \left. \frac{d^{2}\sigma_{0}}{dE_{b} d\Omega_{b}}
\right|_{\theta_b =0} =
\frac{m_{a} m_{b} m_{A} m_{B}}{(2\pi)^{5}\, 2\,\sqrt{s}}
\frac{p_{b}}{p_{a}} \int d\Omega^{d}_{\pi}\,
\frac{p^{d}_{\pi}}{s_{d}}\,\overline{|T_{0}|^{2}}\, .
\label{incl0}
\end{eqnarray}

\vspace{0.1in}
 Fig.~2(a) and 2(b) show the final results of $\sigma$ and
 $R \equiv \sigma_{0}/\sigma$. They are compared with the experimental
data.
The solid lines are our final results including both PDP and s-wave
scattering, while
the dotted line shown in Fig.~2(b) represents the result obtained when
only the contribution from PDP is taken into account.
Comparing the two theoretical cross sections in
Fig.~2(b), it can be seen that the PDP dominates $\sigma_{0}$.  The
contribution from the s-wave rescattering process to $\sigma_{0}$ is
thus small, though
it helps to improve the fit of the calculated final $\sigma_{0}$ to
the experimental data, particularly at the off-resonance region.
The inclusive cross section data
$\sigma^{exp}$ are taken from Ref.~\cite{Prou},
while the experimental $R$ ($R^{exp}$) are obtained by  $D_{xx}$ and
$D_{zz}$ of Refs.~\cite{Glas2,Prou} and $\sigma^{exp}$ of Ref.~\cite{Prou}.
As seen  in the Fig.~2(a), $\sigma^{exp}$ is reproduced very well by
the calculation.
In the resonance region, the $R^{exp}$-values are rather small;
$R^{exp} \approx 0.025$, implying that $\sigma^{exp}$
contributes only about 2.5\% to the
total exclusive cross section. However, $R^{exp}$ becomes larger
at both tail regions of the resonance.

\vspace{0.1in}
 The good fit of the calculated $R$ to the data seems to support
strongly that the observed $\sigma_{0}$ indeed comes from PDP.
This conclusion is further supported by the data of $R$ for nuclear
targets available for the $d$, $^{12}$C, $^{40}$Ca and $^{208}$Pb
targets.
For the deuteron target, the $R^{exp}$-values are larger
by about a factor of 2$\sim$4 as compared with those of the proton target.
The $R$-values for other nuclear target are about the same as those
of the deuteron target. The observed increase of the $R$-values
may easily be understood if one assumes that $\sigma_{0}$ comes from PDP.
Since the dominant part of $\sigma$ comes from TDP,
$\sigma$ for the deuteron target is expected to be about 4/3 of
$\,\sigma\,$  for the proton target due to isospin, while $\sigma_{0}$
of the deutron should be  about 4 times of $\sigma_{0}$ for the proton
target. Thus, it is expected
that the $R$-values may become larger by about a factor of 3 for the
deutron target case, as compared with the proton target case. This
agrees very well with the experimental factor 2$\sim$4.
Since the ratio of the number of protons to that of  neutrons
contributing to the reaction may roughly stay to be unity, the
$R$-value for the heavy nuclei
should roughly be equal to that of the deuteron target case,
which also agrees with the observation.

Finally, we remark that $\sigma_{0}$ may come from the
TDP via the  $\Delta S=0$ interaction term involved in the
$\hat{t}_{NN,\Delta N}$. Such a term has recently been determined
from the analysis of the $p(p,n)\Delta^{++}$ reaction data~\cite{Ray}.
Using the $\hat{t}_{NN, \Delta N}$ operator determined in
Ref.~\cite{Ray}, one can estimate $\sigma_{0}\,$. It has been
found that both magnitude and energy dependence of $\sigma_{0}$
thus estimated do not fit the data very well; the magnitude is
larger by about a factor of 2 than $R^{exp}$, and also the
$\omega$-dependence is quite different from what is observed.
This might have been caused by the fact that the analysis made in
Ref.~\cite{Ray} is done without taking into account PDP.

In summary, we have shown that $\sigma_{0}\;$, deduced from the
data of the spin transfer data $D_{xx}$ and $D_{zz}$ together with the
inclusive cross section $\sigma\;$, can be well explained by the
calculations
that take into account PDP and the s-wave rescattering effects.

\acknowledgments

We would like to express our sincere thanks to Professor T. Udagawa for
his careful reading of the manuscript and valuable comments.
This work is supported in part by the U. S. Department of Energy
under contract DE-FG03-93ER40785.

\begin{figure}
\caption{\label{fig1}Feynman diagrams for $p(p,n)N\pi\,$
reaction. Figures (a) and (b) show the p-wave interacton($\Delta$
excitations) in the target and projectile respectively, while figures (c)
and (d) show s-wave rescatterings in the target and projectile.}
\end{figure}
\begin{figure}
\caption{\label{fig2}Zero degree neutron spectra for the
reaction $p(p,n)N\pi\,$
at $E_{p}=795$ MeV. The inclusive cross section (a) and
the ratio $R=\sigma_{0}/\sigma$ (b) are shown. See text for details.
}
\end{figure}

\end{document}